# Development of a novel transposed cable with in-plane bending of HTS tapes (X-cable)

**Juan Wang[1], Rui Kang[1], Tao Xing[2], Xin Chen[1], chao yang[1], Yingzhe Wang[1], Xiongzhuang Li[2], and Qingjin Xu[1]***

1 Key Laboratory of Particle Acceleration Physics & Technology, Institute of High Energy Physics,
Chinese Academy of Sciences, Beijing 100049, China

2 Baoding Tianwei Wire Manufacturing Co. LTD, Baoding 071023, China

E-mail: xuqj@ihep.ac.cn

## Abstract

A compact HTS cable that is able to carry large current density is crucial for developing high field accelerator magnets. We are reporting a novel HTS cable (named X-cable) that could achieve a high current density as the Roebel cable, but is implemented by in-plane bending stacked HTS tapes directly to realize the transposition. The cable is jointly developed with an industrial company on a production line: ready for large scale production from the beginning. Recently, a prototype cable with REBCO coated conductor has been successfully fabricated. Test results show no significant degradation, demonstrated the feasibility of such cable concept. In this paper, the cable's design concept, in-plane bending performance of the REBCO tapes, fabrication procedure and test results of this first prototype cable will be presented.

## I. Introduction

Many HTS materials have been commercially supplied presently, such as $Bi_2Sr_2Ca_2Cu_3O_x$ (Bi2223) tapes [1-3], $Bi_2Sr_2CaCu_2O_x$ (Bi2212) wires [4], $RE-Ba_2Cu_3O_{7-\delta}$ (REBCO) [5-8] coated conductors and so on. Among them, REBCO coated conductor is one of the promising candidates to be employed in high field applications, provided with high current-carrying capacity at high magnetic field, mechanical stability and no need for heat treatment during the coil fabrication process. For accelerator magnets, it is almost mandatory to wind the coils with a compact large-current cable that contains tens or even hundreds of conductors to reduce inductance and complexity of the coils. However, HTS conductors like REBCO have a large aspect ratio, in addition to the brittle nature of the ceramic material. All these characteristics have brought huge challenges to the magnet applications.

In the past few years, a large number of HTS cable concepts have been proposed. The simplest one is just to stack few tapes, which is however fully non-transposed and is impossible to achieve very large current [9,10]. To obtain higher current, by arranging few stacks together with different methods, one get the so called "Quasi-Isotropic Strand" or the "STARS conductor", which are also not transposed [11]. Takayasu et al firstly introduced the twisted stacked-tape conductor concept [12]. After this, few more partially transposed cable concepts are introduced [13-15], which however all have a rather low current density and large critical bending radius. The conductor on round core (CORC) cable design, which is implemented by wrapping several layers of REBCO tapes around a copper core, is another attractive concept [6,16-18]. Such cable is not transposed and can achieve a moderate high current density. The only fully transposed REBCO cable is Roebel cable, which can achieve a current density almost as high as a single tape. However, the high performance of Roeble cable is built on the sacrifice of half of the raw REBCO material, so the cost is at least doubled comparing with the REBCO tapes.

In view of above, we are developing a novel high-current cable for HTS tapes to be used in magnet applications, as shown in figure 1. The cable has a similar structure to Roebel cable, however, in contrast with Roebel cable, which requires punching and assembling of tapes, the cable proposed here is manufactured by in-plane bending stacked HTS tapes directly to realize the transposition, therefore, the cable has a current-carrying capacity as high as Roebel but with a relatively low cost. The design concept and parameters of this cable are firstly introduced in this paper, then the results of in-plane bending tests of ReBCO conductors are reported, which guides the cable design and fabrication parameters. At last, the fabrication process and test results of the first prototype cable are presented. Comparing with the single ReBCO tapes, the cable shows no degradation, demonstrated feasibility of this concept.

## II. Design concept of the cable

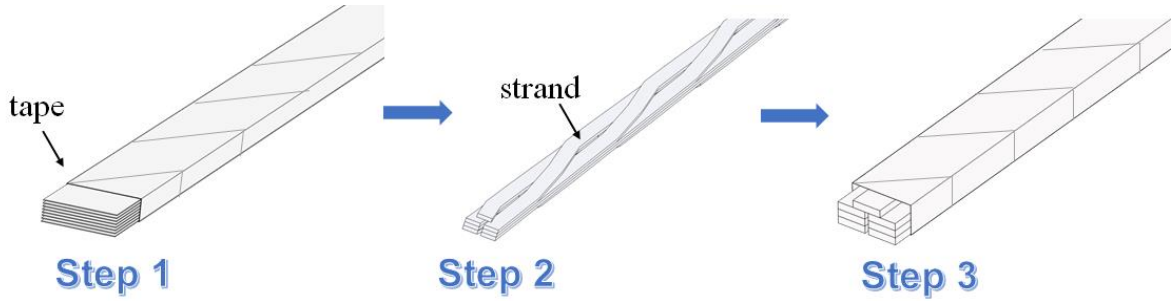

Figure 1. Schematic illustration of the X-cable and the fabrication steps.

Figure 1 illustrates the design concept and fabrication steps of the cable. The entire process can be divided into three steps. In step 1, several REBCO tapes are stacked and wrapped by thin copper tapes, making the so-called strand. Then several strands (9 in this prototype) are in-plane bent and cabled together, as shown in step 2. Finally, the cable is wrapped with a certain material to avoid spring back of the strand.

We could represent the geometry of a strand in the transposed cable by the following parameters, as shown in Figure 2: transposition section length $l_{flat}$, straight section length $l_{side}$, cable width $w_{cable}$, strand width $w_{strand}$, space between two strands $w_{air}$, cable thickness $t_{cable}$, strand thickness $t_{strand}$, side angle $\alpha$, transposition period length L, which is equal to number of strands multiplied by $l_{flat}$, and the radius of the in-plane bending $R_{transition,}$ determined by the in-plane bending performance of the HTS tapes.

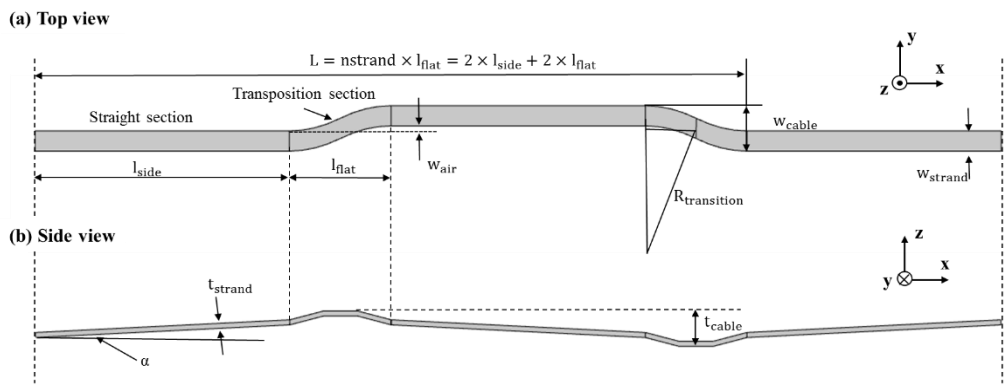

Figure 2. The schematic sketch of one strand in the X-cable, (a) top view; (b) side view.

## III. In-plane bending performance of the REBCO conductors

The cabling process involves in-plane bending of the ReBCO tapes. To find out an optimized bending radius, we have started the investigation of the in-plane bending effect on the current-carrying capacity $I_c$ of REBCO tapes from 2017. $I_c$ measurements of REBCO tapes with different in-plane bending radii have been carried out for many times. Before bending, several tapes are stacked and wrapped with copper foil as one sample, to simulate the strand in the cable. The measurement tool is shown in figure 3(a). For comparison, we tested REBCO tapes from two different manufacturers using the same measurement method, as shown in Table 1. The critical bending radius is defined with 5% $I_c$ degradation of the sample. Note that each sample is only bent to a specific radius. We test each sample five times, and each test is performed after the sample naturally returns to room temperature and cool down again to superconducting state. Figure 3(b) presents the typical test results of in-plane bending performance for REBCO conductors from two different manufacturers. For SuAM tapes, the critical bending radius is about 400 mm, whereas for SST tapes, it is 500 mm.

Table 1. Characteristics of REBCO tapes used in in-plane bending experiment

| REBCO tapes | | |
|---|---|---|
| Manufacture | SST* | SuAM** |
| Width and thickness | 40 mm × 90 um | 40 mm × 80 um |
| Critical current at self-field (1uV cm$^{-1}$) | ≥170 A | ≥105 A |
| The thickness of the silver cap layer | ~1.5 um | ~2 um |
| The thickness of the superconducting layer | ~1 um | ~1 um |
| The thickness of the Hastelloy substrate layer | ~50 um | ~60 um |
| The thickness of the copper stabilization layer | ~18 um | ~5 um |

*SST is the abbreviation of Shanghai Superconducting Technology Co., Ltd., Shanghai, China.
**SuAM is the abbreviation of Suzhou Advanced Materials Research Institute, SuZhou, China.

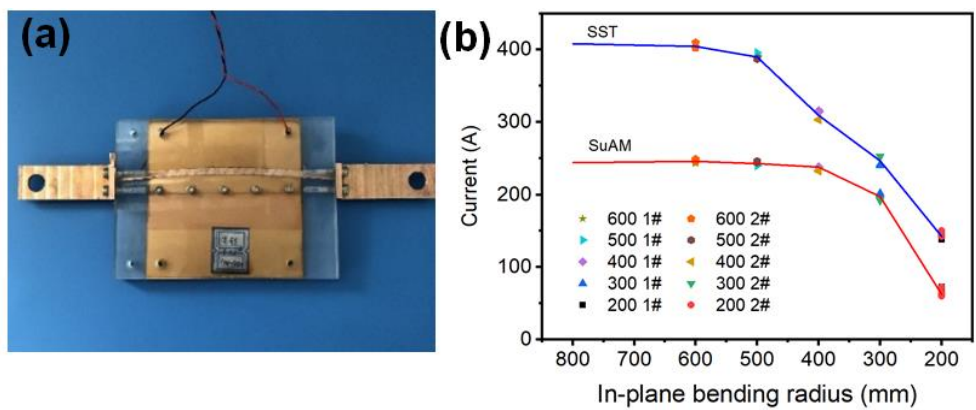

Figure 3. (a) Measurement tool for in-plane bending performance. (b) Critical current at different in-plane bending radius.

## IV. Cable fabrication and test results

A. cable fabrication

In July 2020, after a detailed study of the in-plane bending performance of REBCO tapes, IHEP signed a cooperation agreement with Baoding Tianwei Wire Manufacturing Co. LTD, to develop the proposed cable directly on a production line in the company. In May 2021, we successfully manufactured the 1$^{st}$ prototype cable. The equipment for fabricating and transposing strands is shown in figure 4 (a) and (b). The original cable design consists of nine strands, and each strand consists of eight tapes. However, only twenty REBCO tapes are distributed in different patterns of four strands in the prototype cable, the remained places all replaced by Hastelloy tapes, as shown as in Figure 4 (c). In such a way, this prototype can represent well a fully cable with a moderate cost. The transposition period length L (1.6 m) is half of the first Bi-2223 Roebel cable fabricated by Siemens Corporate Technology [3]. The detail parameters of the cable are listed in table 2.

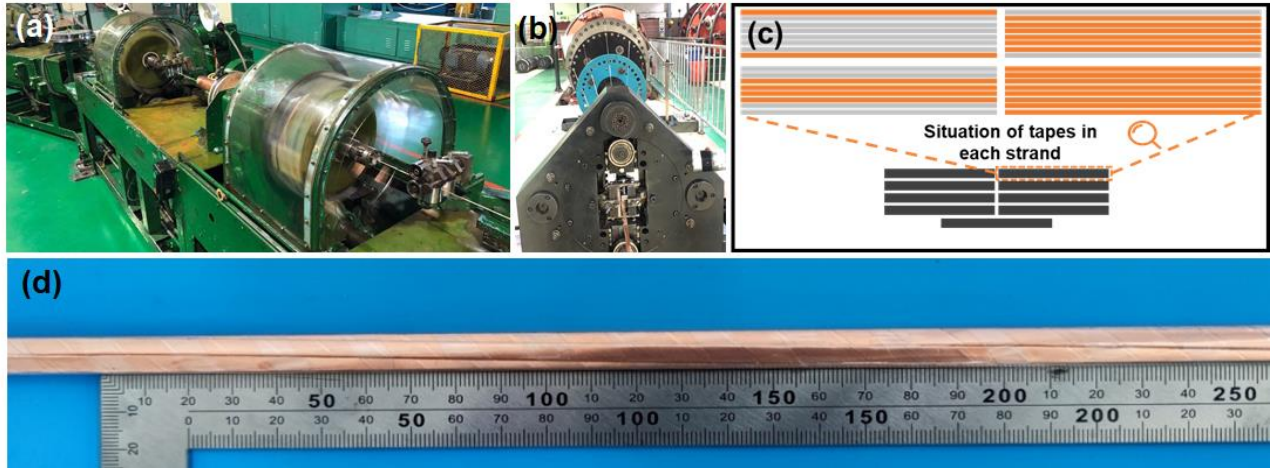

Figure 4. (a) Equipment for strand fabrication. (b) Equipment for transposition and cabling. (c) Cross-section of the prototype cable: orange representing REBCO tapes, gray representing Hastelloy tapes. (d) Picture of the prototype cable showing the in-plane-bending section.

Table 2. parameters of the prototype cable

| Prototype cable | |
| --- | --- |
| Cable length | ~10 m |
| Cable width | ~9.76 mm |
| Cable thickness | ~ 7.4 mm |
| transposition section length $l_{flat}$ | 180 mm |
| Number of strands | 9 |
| transposition period length L | 1.6 m |

B. Test results

The total length of the prototype cable is about 10 m (Schematic diagram is shown in figure 5). To transport the cable from the company to the test lab, we wind the cable in a disc with a diameter of 840 mm. Before winding, we also cut two 1.2 m samples (sample 1 and sample 2) for comparison. In these two samples, all the strands contain an in-plane bending section at least once, where degradation most likely happens. During the cabling process, a damage occurred at the position of 4.5 m due to a local defect in one strand (the copper foil broke because of large tension and was manually repaired), significant crease can be observed from the surface, so we cut the sample near by the damage to investigate the reason, named sample 3 (1.77 m). At the same time, we cut a 1.8 m sample (sample 4) contained a complete transposition length next to the sample 3 from the prototype cable which is wound in the disc.

The measured $I_c$ of all REBCO tapes in sample 1, sample 2, sample 4 and most of the tapes in sample 3 are comparable to the $I_c$ of the short single tapes from which they are composed (Figure 6), except tapes P3-S6T1, P3-S6T2, P3-S6T3, P3-S6T4 and P3-S6T8 in sample 3, due to the defect mentioned above. The results show no performance degradation of the tapes if there is no visible damage occurred during the whole cabling process.

Figure 7 shows the damage of sample 3. In order to analyze the cause of the damage, we disassembled sample 3 and found that the copper tapes outside S6 at the position of the damage is thicker than normal, due to the breakage and manual repair of the copper foil. The difference in thickness along the length caused damage of that strand during the cabling process. Higher strength copper foil would be preferred in future to avoid such accidents for long length production.

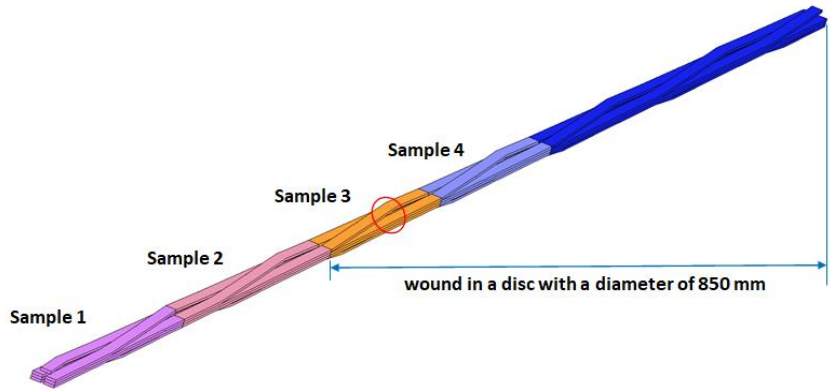

Figure 5. Schematic illustrate of the prototype cable (10 m). The damage is at the location of red circle.

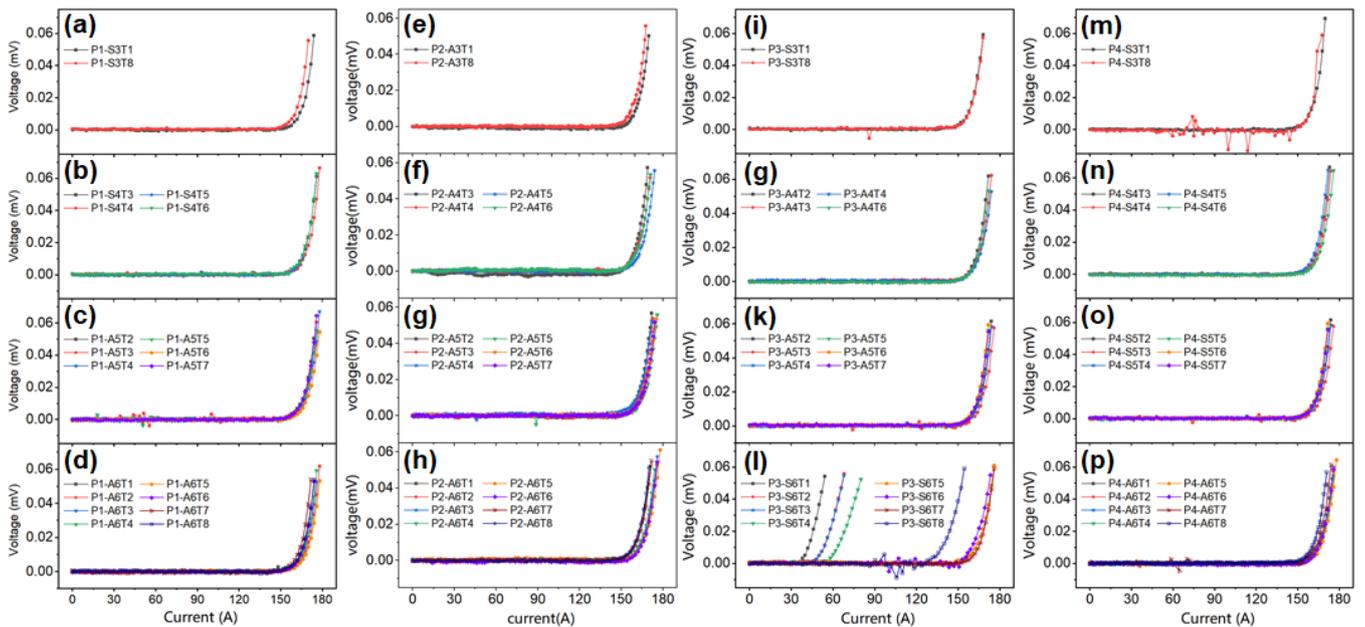

Figure 6. V-I curves of tapes in (a-d) Sample 1, (e-h) Sample 2, (i-l) Sample 3, (m-p) Sample 4 after transposing. P is for sample. S is for strand. T is for tape.

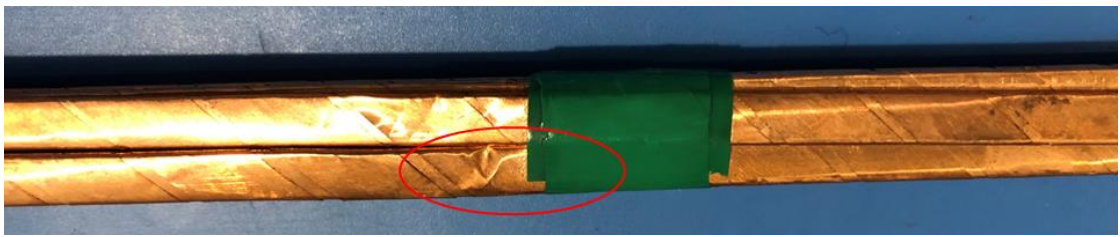

Figure 7. The defect of sample 3 of the prototype cable.

## VI. Conclusion

In this work, we proposed a novel HTS cable (named X-Cable) and successfully fabricated a prototype with REBCO coated conductors, which is implemented by in-plane bending HTS tapes directly to realize the transposition. The advantages of the cable are: (1) compact; (2) high current-carrying capacity; (3) modest cost, and (4) jointly developed with industrial factory at the very beginning to avoid possible modifications during the technology transfer from laboratory to factory. The test results show that the whole fabrication process doesn't degrade the current-currying performance of the REBCO tapes, confirms the feasibility of this novel HTS cable. The cable is very promising for applications in high field magnets. Optimization of the design parameters, fabrication process and detailed performance study of this cable will be carried out in future.

## Acknowledgement

The work is supported by the Strategic Priority Research Program of the Chinese Academy of Sciences (Grant No. XDB25000000) and the National Key R&D Program of China (Grant No. 2018YFA0704200).